\documentclass[aps,prx,twocolumn,10pt,english,superscriptaddress,floatfix]{revtex4-2}
\usepackage[english]{babel}
\usepackage[utf8]{inputenc}
\usepackage{graphicx}
\usepackage{dcolumn}
\usepackage{bm}
\usepackage{amsmath}
\usepackage{natbib}
\usepackage{color}
\usepackage{amssymb}
\usepackage{palatino}
\usepackage[hidelinks]{hyperref}

\begin{document}

\title{Satellite-assisted quantum communication with single photon sources and atomic memories}

\author{V. Domínguez Tubío}
\affiliation{QuTech, Delft University of Technology, 2628 CJ Delft, The Netherlands}
\affiliation{Kavli Institute of Nanoscience, Delft University of Technology, 2628 CJ, Delft, The Netherlands}
\author{M. Badás Aldecocea}
\affiliation{Space Engineering Department, Delft University of Technology, 2629 HS Delft, The Netherlands}
\author{J. van Dam}
\affiliation{QuTech, Delft University of Technology, 2628 CJ Delft, The Netherlands}
\affiliation{Kavli Institute of Nanoscience, Delft University of Technology, 2628 CJ, Delft, The Netherlands}
\affiliation{Quantum Computer Science, EEMCS, Delft University of Technology, 2628 CJ, Delft, The Netherlands}
\author{A. S. S\o rensen}
\affiliation{Center for Hybrid Quantum Networks (Hy-Q), Niels Bohr Institute, University of Copenhagen, Blegdamsvej 17, Copenhagen DK-2100, Denmark}
\author{J. Borregaard}
\affiliation{QuTech, Delft University of Technology, 2628 CJ Delft, The Netherlands}
\affiliation{Department of Physics, Harvard University, Cambridge, Massachusetts 02138, USA}

\date{\today}

\begin{abstract}
Satellite-based quantum repeaters are a promising means to reach global distances in quantum networking due to the polynomial decrease of optical transmission with distance in free space, in contrast to the exponential decrease in optical fibers. We propose a satellite-based quantum repeater architecture with trapped individual atomic qubits, which can serve both as quantum memories and true single photon sources. This hardware allows for nearly deterministic Bell measurements and exhibits long coherence times without the need for costly cryogenic technology in space. We develop a detailed analytical model of the repeater, which includes the main imperfections of the quantum hardware and the optical link, allowing us to estimate that high-rate and high-fidelity entanglement distribution can be achieved over inter-continental distances. In particular, we find that high fidelity entanglement distribution over thousands of kilometres at a rate of 100 Hz can be achieved with orders of magnitude fewer memory modes than conventional architectures based on optical Bell state measurements. 
\end{abstract}

\maketitle

\section{Introduction}
The implementation of a quantum internet opens a range of new opportunities for secure communication~\cite{BENNETT20147,Ekert_91,Pirandola2020}, enhanced sensing networks~\cite{Komar2014,Guo2020,Liu2021}, and distributed quantum computing~\cite{buhrman_distributed_2003}. To exploit these opportunities in applications such as protecting and optimizing large-scale power distribution networks~\cite{Tang2023} or probing fundamental constants and geodesy~\cite{Nichol2022}, it is necessary to extend the range of quantum networks to distances of thousands of kilometres.

To carry out quantum communication, we need to use photons to transmit quantum information. However, transmission loss of an optical quantum signal cannot be compensated with standard classical amplification techniques due to the quantum no-cloning theorem~\cite{Wootters1982}. Instead, quantum repeaters have been proposed, where the total distance is divided into smaller segments over which direct transmission is feasible. The segments are then combined either through quantum teleportation~\cite{Duan2001,Sangouard2011} or quantum error correction~\cite{Muralidharan2014,Munro2012} at the repeater nodes to enable faithful transmission over the total distance. 

For optical fiber-based quantum repeaters, the transmission between the repeater nodes decreases exponentially with distance and hundreds of repeater nodes are required to cover distances at the continental scale. Alternatively, fiber-based connections can be replaced with satellite-assisted free-space optical links where transmission decreases only polynomially with the distance~\cite{forges2023,Liorni_2021,gundogan2021,Wallnofer_2022,Khatri_2021,Boone_2015,Gustavo_paper}. For continental scales, this can reduce the required number of repeater nodes by orders of magnitude making up for the arguably higher cost of space-based quantum repeater nodes. 

Quantum key distribution (QKD) has already been demonstrated with the Micius satellite over 1200 km~\cite{yin_entanglement-based_2020}. Reaching larger distances by direct transmission, however, requires a high-orbit satellite due to the limitation from the line of sight of the ground receivers. This substantially increases the transmission loss, making high-rate quantum communication extremely challenging. Alternatively, multiple low-orbit satellites can operate in a quantum repeater architecture to efficiently compensate for the transmission loss and ensure line of sight between distant locations~\cite{Liorni_2021,Wallnofer_2022,Gundogan_2021}. 

Previous theoretical work on satellite-based quantum repeaters assumes the use of probabilistic optical Bell state measurements (BSMs)~\cite{Liorni_2021},  which requires multiplexing of thousands of memory modes in order to reach high-rate communication over global distances. Additionally, the availability of on-demand entangled photon pair sources are often assumed~\cite{Wallnofer_2022,Liorni_2021,Khatri_2021,Boone_2015}, which remains an outstanding technological challenge~\cite{Anwar2022}. Promising candidate systems are solid-state semiconductor quantum dots~\cite{Schimpf2021}, which require cryogenic temperatures to function or multiplexing of SPDC sources~\cite{Chen2023} and single photon sources~\cite{Zhang2008}. 

We propose a satellite-assisted quantum repeater protocol based on trapped individual atomic qubits. Individually trapped Alkali atoms can both function as efficient single-photon sources~\cite{Kimble_2004} and atomic memories due to their long coherence times~\cite{Bluvstein_2022, Young_2020}. Furthermore, they enable nearly deterministic Bell state measurements through Rydberg-mediated two-atom gates~\cite{Levine_2019,Fu_2022,Evered_2023} and the possibility of scaling to hundreds of qubits per repeater node~\cite{Bernien_2021,Covey2023}. Additionally, laser-cooling of the atoms is sufficient to ensure long coherence times, which avoids the need for costly cryogenic technology in space. 

We develop a detailed analytical model of the repeater protocol that considers the main imperfections of both the quantum hardware and the optical link budget. In contrast to Monte-Carlo-based simulations~\cite{Wallnofer_2022}, the analytical model allows us to efficiently simulate long chains of quantum repeater nodes. We show that, for realistic satellite parameters and quantum hardware errors, we can get high-fidelity ($\geq 0.9$) entanglement distribution rates of 100 Hz over continental distances of up to 1500 km using 5 satellites with less than 200 quantum memory modes per satellite. 

\section{Results}
\subsection{Architecture}
\begin{figure*}[htp]
   \centering
    \includegraphics[width=0.9\linewidth]{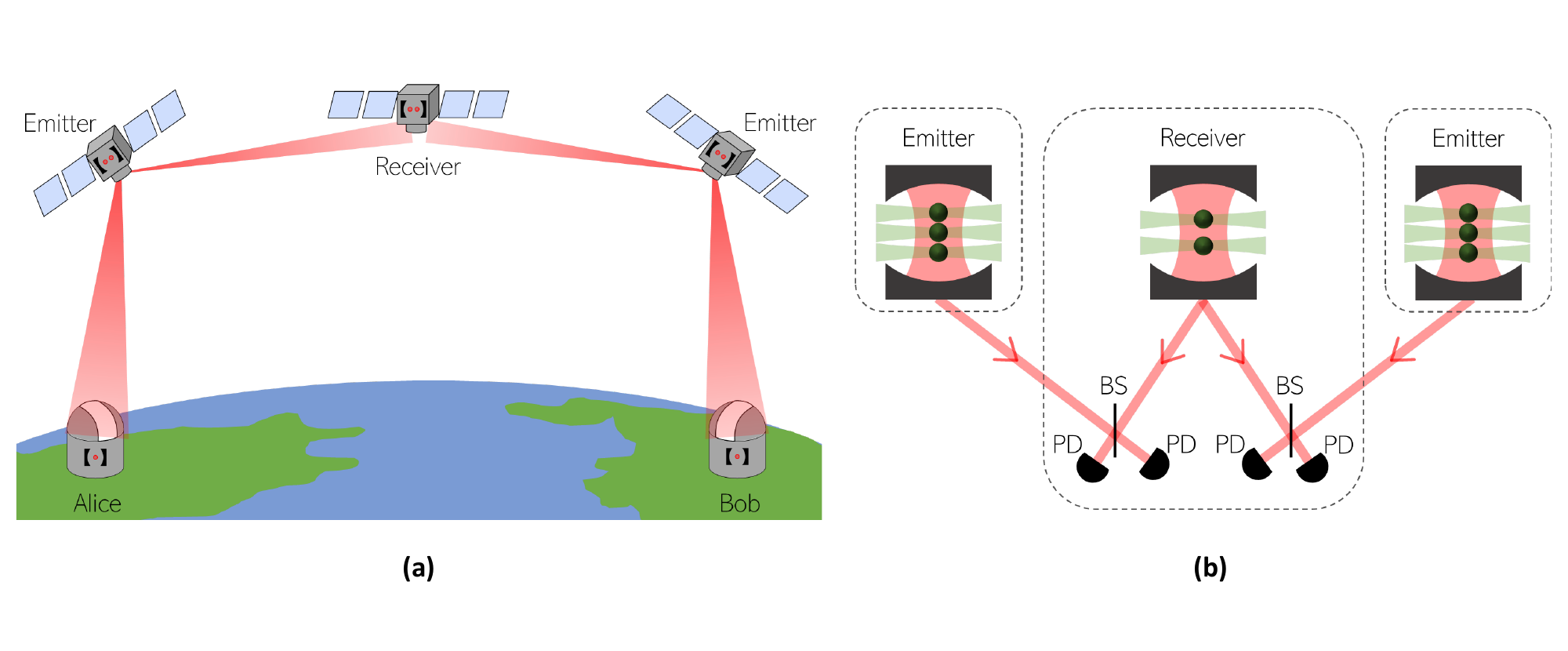}
    \caption{\textbf{Experimental setup. (a)} We consider a down-link scenario where the photons are transmitted from the end-satellites to the ground stations. Each of the satellites contains quantum memories and entanglement swaps are performed on the satellites to link the segments of the repeater (satellite-satellite or satellite-ground stations) once entanglement has been established. \textbf{(b)} Probabilistic BSM. The atom-photon entanglement generated through the emission of a photon from an atom is swapped to atom-atom entanglement within a receiver satellite by means of a linear optics Bell state measurement. This is done using a 50/50 beam splitter (BS) and single photon detectors.}
    \label{fig:fig_1}
\end{figure*}

We consider a down-link scenario where photons are sent from the end-satellites to the ground stations, which is more robust to atmospheric turbulence than the reverse up-link scenario~\cite{andrews_laser_2005,Gundogan_2021}. A high-level sketch of a 3-node repeater is shown in Fig.~\ref{fig:fig_1}a). The quantum repeater consists of different types of satellites. Some satellites act as emitters, sending photons to their neighbouring satellites or to the ground station. Others are receivers, that collect the photons sent by the emitter satellites. Both types of satellites are equipped with atomic quantum memories and act as quantum repeater nodes, as shown in Fig.~\ref{fig:fig_1}b). 

To distribute entanglement between the ground stations, atom-photon entanglement is generated at the repeater stations through pulsed excitation of the atoms. The generated photons are sent from the emitter nodes to the receiver ones, where the atom-photon entanglement is swapped to atom-atom entanglement by means of a linear optics Bell State Measurement (BSM). Once entanglement between two neighbouring elementary links has been successfully heralded, a nearly deterministic entanglement swap operation is performed through Rydberg-mediated atom-atom interactions. If the entanglement generation is successful in all elementary links, the two ground states will share an entangled pair. 

\subsection{Entanglement Generation}

For the generation of atom-atom entanglement, we consider a `two-click' protocol~\cite{Barrett2005}, where each atom is entangled with a time-bin encoded single photon through pulsed excitation. This is then swapped to atom-atom entanglement by a linear optics Bell measurement. We focus on this scheme since it only requires phase stability on the time-scale of the time-bin separation and does not have the fundamental trade-off between rate and fidelity of `single-click' protocols~\cite{Cabrillo1999}. Furthermore, we consider a scenario where the entanglement swapping happens at a heralding station placed on the receiving satellite. The loss will thus only affect one of the photons and the usual advantage of the single-click protocol of having a higher rate for long distances is thus not applicable in this scenario.

We now describe the steps of the entanglement generation in more detail. Fig.~\ref{fig:fig_2}a) shows how individual atoms are trapped with optical tweezers inside macroscopic, near-concentric optical cavities~\cite{Bernien_2021,Deist2022,Covey2023}. Throughout this paper, we will focus on an implementation with Rubidium (Rb) atoms though other Alkali atoms such as Cesium could also be used. The closed optical transition between  the $5\text{S}_{1/2}$,${|\text{F}=2,\text{m}_{\text{F}}=2\rangle}$ ground state and the $5\text{P}_{3/2}$, ${|\text{F}'=3,\text{m}_{\text{F}'}=3\rangle}$ excited state in $^{87}$Rb allows for spin-photon entanglement through pulsed excitation. For simplicity, we first describe this process for a single atom and then discuss how to perform the operation for a collection of atoms. 

First, the atom is prepared in a superposition of the spin states,
\begin{equation}
    |\phi\rangle = \frac{1}{\sqrt{2}}\left(|0\rangle + |1\rangle\right).
    \label{eq:equation_1}
\end{equation}
by means of standard optical pumping and two-photon Raman driving~\cite{Bluvstein2022,Levine2022}. We imagine that $|0\rangle=|\text{F}=1,\text{m}_\text{F}=1\rangle$ and $|1\rangle=|\text{F}=2,\text{m}_\text{F}=2\rangle$ in the $5^2\text{S}_{1/2}$ ground state manifold. Next, a short optical $\pi$-pulse is applied to induce the transition, $|1\rangle\rightarrow|ex\rangle$, where $|ex\rangle=|\text{F}'=3,\text{m}_\text{F}'=3\rangle$ in the excited $5^2\text{P}_{3/2}$ manifold. The excited state will subsequently decay back to $|1\rangle$ by emission of an early cavity photon, $|e\rangle$. Next, two-photon Raman driving is used to flip the population of the ground states i.e. $|0\rangle \leftrightarrow|1\rangle$ after which a second optical $\pi$-pulse is applied resulting in the emission of a late photon, $|l\rangle$ if the atom is in the $|1\rangle$ state. Ideally, this procedure results in the spin-photon entangled state
\begin{equation}
    |\psi\rangle_{\text{sp}} = \frac{1}{\sqrt{2}}\left(|0\rangle|e\rangle + |1\rangle|l\rangle\right).
    \label{eq:equation_2}
\end{equation}
When performing this operation on a collection of atoms it is important to prevent that the emission from one atom interferes with another. To this end, we propose to operate in a sequential manner where all atoms are initially prepared in the state $(|0\rangle+|1\rangle)/\sqrt{2}$ and then addressed sequentially with the optical $\pi$-pulse for generation of atom-photon entanglement. An additional laser at 1530 nm addressing the transition between the $5^2\text{P}_{3/2}$ and $4^2\text{D}_{3/2}$ excited manifolds is applied to all atoms except the one subject to the optical $\pi$ pulse. Specifically, a $\pi$-polarized laser will couple the excited $|ex\rangle$ to the $|ex'\rangle=|\text{F}''=3,\text{m}_{\text{F}''}=3\rangle$ hyper-fine level of the $4^2\text{D}_{3/2}$ manifold. This is done to effectively shift the $|1\rangle\leftrightarrow|ex\rangle$ transition out of resonance for the other atoms since the excited dressed states will be detuned from the cavity resonance by $\pm\hbar\Omega$, where $\Omega$ is the Rabi frequency of the laser-driven $|ex\rangle\leftrightarrow|ex'\rangle$ transition~\cite{li2024}. Having $\Omega\gg g\sqrt{N}$, where $g$ is the single photon Rabi frequency of the cavity coupled $|1\rangle\leftrightarrow|ex\rangle$ transition ensures that there is effectively no coupling of the atoms to the cavity field. In this way, atom-photon entanglement can be attempted with each atom sequentially as described above.   

The photons collected from the cavity are transmitted to either one of the ground stations or another satellite depending on the specific location of the satellite in the repeater chain. In both cases, a linear optics BSM is performed at the destination to herald atom-atom entanglement, as shown in Fig.~\ref{fig:fig_1}b). The latter is carried out with a $50/50$ beam splitter and single photon detectors following the scheme of Ref.~\cite{Barrett2005}. If the BSM is successful, meaning we have measured an early and a late photon, we have accomplished entanglement in an elementary link, i.e. satellite-satellite or satellite-ground entanglement,

\begin{equation}
    |\psi\rangle_{\text{sp}}\otimes |\psi\rangle_{\text{sp}} \xrightarrow{BSM} |\psi\rangle = \frac{1}{\sqrt{2}}\left(|01\rangle \pm |01\rangle\right).
    \label{eq:equation_3}
\end{equation}
The phase of the superposition is determined by which detectors record the photons. Since this information is known, it is possible to change the phase with local qubit operations, if necessary.   

\subsection{Entanglement SWAP}

Following the successful atom-atom entanglement generation of neighboring links in the setup, we carry out a SWAP between the entangled atoms of the different links. Previous satellite-based quantum repeater schemes have considered photonic Bell measurements similar to the entanglement generation step~\cite{Liorni_2021,gundogan2021,Boone_2015} to achieve this. However, the downside of performing a photonic BSM is that it has an intrinsic failure probability of 50\%, which has a detrimental effect on the overall rate of the repeater. To circumvent this, we consider nearly-deterministic Bell measurement between pair of atoms through the Rydberg interaction. To do so, the atoms are spatially re-arranged. This can be performed on a timescale of milliseconds~\cite{Bernien_2021,Bluvstein2022}, which is on the order of the communication time between segments for the parameters considered below. This is different from the situation in the entanglement generation step where the typical time scale of the local operations is on the microsecond scale for typical experimental parameters~\cite{Bernien_2021,Covey2023} and thus negligible compared to the communication time. 

\begin{figure*}[htp]
    \centering
    \includegraphics[width=0.90\linewidth]{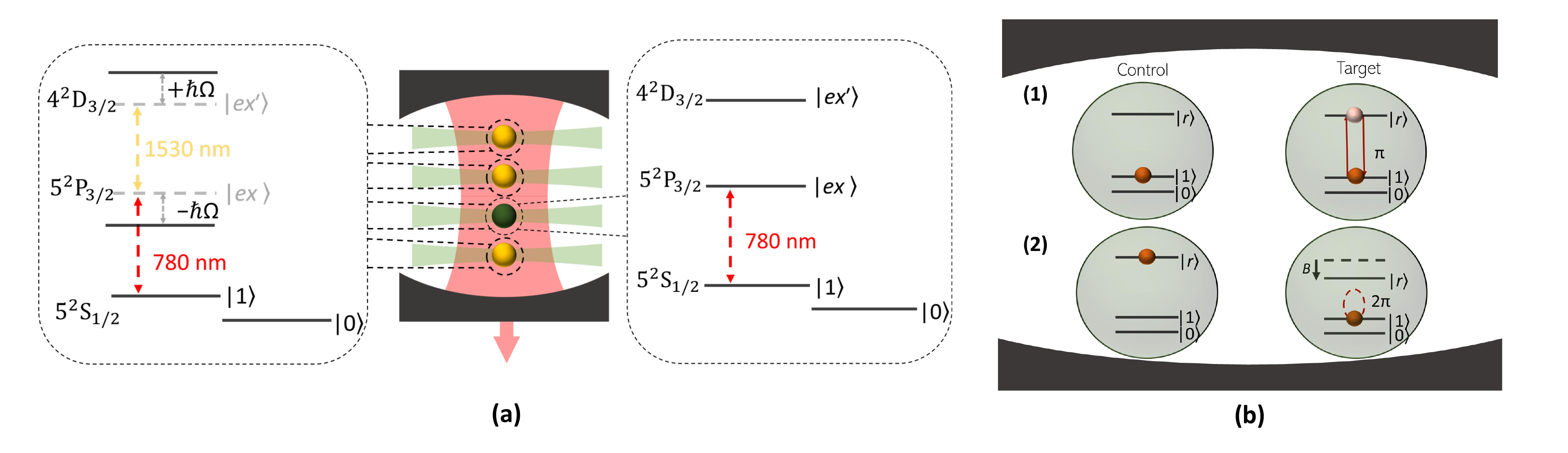}
    \caption{\textbf{Individually trapped atoms in a cavity as the main hardware. (a)} The quantum memories chosen are Rubidium (Rb) atoms in a cavity, which also work as emitters. With a laser resonant with the $|1\rangle \rightarrow|ex\rangle$ transition, the atom is excited to the $|ex\rangle$ state from which it will decay back to the $|1\rangle$ state with the emission of a photon. To selectively address only one atom in an entangling attempt, the other atoms are shifted out of resonance through strong driving of the $|ex\rangle \rightarrow |ex'\rangle$ transition. \textbf{(b)} Nearly deterministic BSM. After the elementary links are entangled a nearly deterministic Bell state measurement is carried out to distribute entanglement between Alice and Bob. The latter is performed by applying a CNOT gate between the atoms and a Hadamard gate in the control atom. To perform the 2-qubit gate, we exploit the well-known Rydberg blockade: The $|1\rangle\to|r\rangle$ transition of the target atom is resonant with our driving field if (1) the control atom is in state $|0\rangle$ but (2) shifted out of resonance by the Rydberg interaction if the control atom is in state $|r\rangle$.    }
    \label{fig:fig_2}
\end{figure*}

A Rydberg-mediated CZ gate between the two atoms~\cite{jaksch2000,levine2019,Jandura2022} is assumed for the Bell measurement. Since the originally proposed schemes in Ref.~\cite{jaksch2000}, there have been a number of further developments to increase the performance of Rydberg-mediated two-atom gates~\cite{levine2019,Jandura2022} leading to experimentally reported gate fidelities 
exceeding $99\%$~\cite{Evered_2023}. To illustrate the basic idea of how the Rydberg-interaction allows for a two-atom controlled phase gate, we focus, however, on one of the original schemes of Ref.~\cite{jaksch2000} illustrated in Fig.~\ref{fig:fig_2}b) for simplicity. First, a $\pi$ pulse is applied to the control atom which makes the transition $|1\rangle\to|r\rangle$, where $|r\rangle$ is a Rydberg state with a high principal quantum number ($n\approx 60$). Then, a $2\pi$ pulse is applied to the target atom. If the control atom is in state $|0\rangle$, this pulse will make the transition $|1\rangle\to|r\rangle\to-|1\rangle$ on the target atom. However, if the control atom is in the Rydberg state, the Rydberg interaction will shift the transition out of resonance such that the target atoms remain essentially unperturbed by the pulse. A final $\pi$ pulse brings the population of the control atom back to $|1\rangle$ and concludes the gate. This operation amounts ideally to a controlled phase gate between the atoms, which is sufficient to perform a Bell state measurement.      

To realize a Bell measurement, it is necessary to apply a Hadmard gate on the target atom before the CZ gate and a Hadamard gate on the control atom after the gate, which can be done with two-photon Raman driving~\cite{Bluvstein_2022,Levine2022}. The qubit states are then measured by first driving the closed $|1\rangle\leftrightarrow|ex\rangle$ transition and collecting the emitted light which detects if the atom is in the $|1\rangle$ state. If no light is detected, the atom is either in the $|0\rangle$ state or it could have been lost from the trap. The latter can result from the re-arrangement step, the two-atom gate or simply from imperfect vacuum. To herald whether the atom was lost, a Raman drive is applied to bring the population from $|0\rangle$ to $|1\rangle$ followed by a second detection. If light is collected, the atom was in the $|0\rangle$ state while if no light was collected, the atom is assumed lost. We note that this means that the Bell measurement is in fact not deterministic since there is a non-zero probability that the atom was lost in which case the operation fails.      

\subsection{Performance}
To evaluate the performance of the repeater protocol we model a number of imperfections at both the quantum hardware level as well as in the optical transmission budget. Below, we describe these imperfections at a high level and refer to Appendix~\ref{app:A} and the supplemental material~\cite{SM} for additional details of the modelling. At the quantum hardware level, we consider different imperfections in the entanglement generation, the quantum memories, and the entanglement swap.

\textit{Entanglement generation}: In the generation of spin-photon entanglement, we consider undesired two-photon emissions and imperfect coupling to the cavity mode. After the entangled photon is sent from the emitter to the receiver, a probabilistic optical BSM is performed. The photons interfering may not be perfectly indistinguishable, which is accounted for by including a non-unity visibility. The efficiency of the detector and the coupling losses from free space to fiber are also included in the link budget (see below). Moreover, dark counts can generate a "click" in the detectors, despite no photon being transmitted, which we include with a non-zero dark count probability.

\textit{Quantum memories}: The atom-atom entanglement will decohere over time due to dephasing of the spin state of the atoms. We model this as single qubit dephasing channels acting on each atomic qubit leading to an exponential decrease of coherence with time. Furthermore, the atoms can get lost from the traps due to imperfect vacuum, which we model as an erasure channel with an exponential decay of the qubit population with time. 

\textit{Entanglement SWAP}: After entanglement between neighboring links is successfully achieved, we carry out the atomic Bell measurements in all the satellites of the chain to enable entanglement swapping to the ground stations. To account for imperfections in the CZ-gate, we assume that the gate succeeds perfectly with some probability $p_{\text{swap}}$ while with probability $1-p_{\text{swap}}$ the swap results in a `garbage' state with zero fidelity with the desired Bell state. In addition, we also include a finite probability that the atoms participating in the Bell measurement are lost, which also destroys the entanglement but is an heralded error.  

\textit{Optical link budget}: We compute free space propagation losses assuming a fundamental Gaussian beam under transmitter pointing jitter. The latter is assumed to be described by a radially varying Rayleigh probability distribution function. Furthermore, in satellite-to-ground links, we include the atmospheric effects of Rayleigh scattering and beam widening due to turbulence. Finally, internal losses in the terminals, due to non-ideal operation of the optical elements (i.e. absorption in lenses and mirrors) are included. 

In our simulations, we consider two different satellites. One has the characteristics of the Micius satellite~\cite{yin_entanglement-based_2020}, namely a telescope with $15$ cm radius and $0.41$ $\mu$rad pointing error. In addition, we consider an improved second satellite, with a 50-centimeter radius telescope. In both cases, the satellite orbit is at a height of $500$ km from the surface of the Earth, the same as the Micius satellite, and both ground stations have a 60-centimeter radius telescope. We also assume the satellites to be in a `string of pearls' configuration following an equatorial orbit. The ground stations are assumed to be at a height of 2000 meters to avoid Mie scattering. Mie scattering is produced by particles of size comparable to the wavelength of light, mainly due to atmospheric aerosols. These particles are more abundant in the lower atmosphere and can be neglected for higher elevations~\cite{brattich_measurements_2019}.

As detailed in the supplemental material~\cite{SM}, we can derive an analytical estimate of the fidelity of final Bell pairs distributed between the ground stations taking into account all of the aforementioned errors and losses. In order to do this, we adopt a model where errors either lead to a completely dephased Bell state of the from $\rho_{\text{deph}}=(|01\rangle\langle 01|+|10\rangle\langle 10|)/2$ or end up in a non-specified `garbage' state $|g\rangle$ with zero overlap with the desired target state $|\psi\rangle=(|01\rangle+|10\rangle)/\sqrt{2}$. In the entanglement generation, two-photon errors, memory dephasing and non-perfect optical visibility results in errors of the first type. We model the errors from dark counts as resulting in a garbage state as a worst case scenario. This allows us to express the density matrix describing the entangled pairs in the elementary links as $\rho_{\text{link}}=\alpha|\psi\rangle\langle \psi| + \beta\rho_{\text{deph}} + \gamma |g\rangle\langle g|$. The dependence of the coefficients $\alpha,\beta$, and $\gamma$ on the physical parameters such as two-photon emission probability, transmission loss, dark counts and memory coherence time are given in the supplemental material~\cite{SM}. 

Finally, we model the imperfect entanglement swap at the repeater nodes as succeeding with probability $p_{\text{swap}}$ resulting in an error free swap while with probability $1-p_{\text{swap}}$, the swap results in a garbage state. This allows us obtain a compact expression of the final entangled state between the end-nodes of the repeater
\begin{eqnarray}
\rho_{\text{AB}} &=& p_{\text{swap}}^{n_{\text{sat}}}\left(A[n_{\text{sat}}]|\psi\rangle\langle\psi| + B[n_{\text{sat}}]\rho_{\text{deph}}\right) +C[n_{\text{sat}}]|g\rangle\langle g| \nonumber \\
&&+ \left(1-p_{\text{swap}}^{n_{\text{sat}}}\right)\left(A[n_{\text{sat}}] + B[n_{\text{sat}}]\right)|g\rangle\langle g|,
    \label{eq:final_state}
\end{eqnarray}
where $p^{n_{\text{sat}}}_{\text{swap}}$ is the probability of no faulty entanglement swap operations across the chain of $n_{\text{sat}}$ satellites. The coefficients $A[n_{\text{sat}}]$, $B[n_{\text{sat}}]$ and $C[n_{\text{sat}}]$ can straightforwardly be found from combining $(n_{\text{sat}}+1)$ states of the form $\rho_{\text{link}}$ (see supplemental material for details~\cite{SM}). Note that, in general, the coefficients $\alpha,\beta$, and $\gamma$ will be different for each of the elementary links. From Eq.~(\ref{eq:final_state}), it follows that the fidelity of the final state with the target Bell state is $F_{\text{AB}} = p^{n_{\text{sat}}}_{\text{swap}}\frac{A[n_{\text{sat}}] + B[n_{\text{sat}}]/2}{A[n_{\text{sat}}]+B[n_{\text{sat}}]+C[n_{\text{sat}}]}$.  

For the computation of the rate, we assume that $N_{\text{mem}}$ photons are transmitted in each elementary link entangling attempt using 2$N_{\text{mem}}$ atoms per emitter satellite since each satellite covers two elementary links. In each attempt, we assume that the $N_{\text{mem}}$ photons are emitted within a time assumed negligible compared to the communication time between the links, which is on the order of milliseconds. We therefore set the repetition time of the entangling attempt to match the longest communication time between links in the repeater chain. 

After each attempt, the entanglement in a link is either swapped if the neighboring link is also successful or discarded before a new attempt is made. We choose this mode of operation because for quantum memories with second long coherence times, as considered in this work, the communication time between elementary links is too long to maintain high fidelity entanglement by storing successful links for multiple entanglement attempts. It is therefore desirable to have enough multiplexing to ensure near-deterministic entanglement generation in the elementary links~\cite{collins2007}. 

To estimate the rate of the repeater, we first consider the probability of generating $n$ Bell pairs in a single elementary link, which is simply 
\begin{equation} 
p_{(g,s)}(n)= \binom{N_{\text{mem}}}{n} p_{\text{link},(g,s)}^n (1-p_{\text{link},(g,s)})^{N_{\text{mem}}-n}
\label{eq:binom}
\end{equation}
where $p_{\text{link},(g,s)}$ is the success probability per photon for ground-satellite ($g$) or satellite-satellite links ($s$). This expression is valid in the regime where the optical transmission is limited by beam divergence rather than pointing errors since the latter can induce correlated errors, whereas the binomial distribution assumes uncorrelated loss.. We verify that this is the case for realistic pointing errors though Monte-Carlo simulations of the distribution of entangled pairs in an elementary link (see supplemental materials for details~\cite{SM}). Since the number of entangled pairs between the ground stations will be limited by the link with the smallest number of successful pairs, we can express the probability of generating $n$ entangled pairs between the ground stations, assuming deterministic swapping, as
\begin{eqnarray}
p(n)&=&\Big(\sum_{i=0}^{n_{\text{sat}}-1}\binom{n_{\text{sat}}-1}{i} p_s(n)^{i}p_s(>n)^{n_{\text{sat}}-(i+1)}\Big) \nonumber \\
&&\cdot \Big(\sum_{j=0}^{2} \binom{2}{j}  p_g(n)^{j}p_g(>n)^{2-j}\Big) \nonumber \\
&&-p_s(>n)^{n_{\text{sat}}-1}p_g(>n)^2,
\end{eqnarray}
where $p_{(g,s)}(>n)=\sum_{i=n+1}^{N_{\text{mem}}}p_{(g,s)}(i)$. 

Since the entanglement swapping is only near-deterministic due to the possibility of atom loss, we need to include the probability of losing some of these pairs due to failed entanglement swaps. This modifies the probability for establishing $n$ entangled pairs between the ground stations to
\begin{equation}
p_{f}(n)=\sum_{i=n}^{N_{\text{mem}}}\binom{i}{n} p(i) (1-p_{\text{loss}})^{n}p_{\text{loss}}^{i-n} ,
\end{equation}
where $p_{\text{loss}}$ is the probability to unsuccessfully combine the entangled pairs in the elementary links to achieve an entangled pair between the ground stations due to loss of at least one atom in any of the elementary link pairs. The final average rate of the repeater is then estimated as
\begin{equation}
    R = \frac{\sum_{i=1}^{N_{\text{mem}}}ip_{f}(i)}{T_{\text{com}}},
    \label{eq:rate}
\end{equation}
where we have assumed that the repetition rate is set by $T_{\text{com}}$ which is the longest, round-trip communication time between elementary links. Additional details about the calculation of the rate and fidelity can be found in the supplementary material where we provide detailed expressions for $p_{\text{link},(g,s)}$, $p_{\text{loss}}$ and $T_{\text{com}}$.  

\begin{figure*}[htp]
    \centering
    \includegraphics[width=0.9\linewidth]{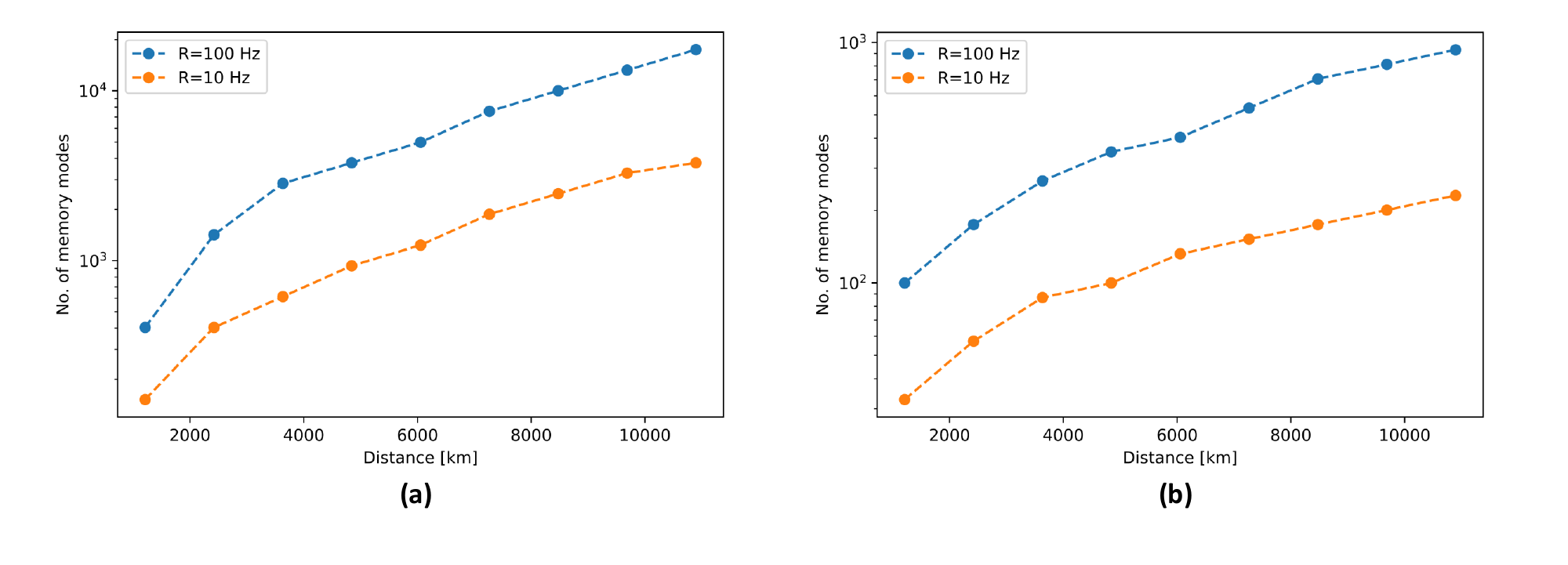}
    \caption{\textbf{Multiplexing capabilities required to reach a rate of 10Hz (orange) or 100Hz (blue) and a fidelity of 0.9 in the entanglement distribution protocol in a satellite chain of 5 satellites. (a)} Number of memory modes needed in a satellite with the characteristics of Micius, i.e. the same size of the telescope ($r=0.15\text{ m}$), pointing error ($\sigma=0.41 \text{ }\mu$\text{rad}) and height ($h=500 \text{ km}$), as a function of the distance between ground stations. \textbf{(b)} Number of memory modes needed in a satellite with a bigger telescope radius than Micius ($r=0.5\text{ m}$) as a function of the distance between ground stations.}
    \label{fig:fig_3}
\end{figure*}

\begin{table*}[htp]
\centering
\begin{tabular}{|c c c|}
\hline
\multicolumn{3}{|c|}{Parameters considered for the simulations} \\
\hline
Prob. of the emitter to emit a single photon & $p_1$ & 0.99\\
Prob. of the emitter to emit two photons & $p_2$ & 0.002 \\
Collection efficiency & $\eta_{\text{coll}}$ & 0.49 \\
Visibility of the photons & $\mathcal{V}$ & 0.999 \\
Prob. of losing an atom in the SWAP & $p_{\text{loss,swap}}$& 0.1\\
Effective loss time in satellite & $T_{\text{loss,sat}}$ & 0.01 s\\
Effective loss time in ground station & $T_{\text{loss,gstat}}$ & 1.5 s \\
Fidelity of the swap & $p_{\text{swap}}$ & 0.995\\
Dark detection probability & $p_{\text{dark}}$ & $10^{-6}$\\
Decoherence time in satellite & $\tau_{\text{c,sat}}$ & 1.5 s\\
Decoherence time in ground station &
$\tau_{\text{c,gstat}}$ & 10 s\\
Photon detection efficiency  & $\eta$ &  0.98\\
\hline
\multicolumn{3}{|c|}{Transmission probabilities in the elementary links} \\
\hline
Trans. prob. sat.-ground. ($r=0.15$m) & $p_{\text{T,sat-ground}}$ & 0.14\\
Trans. prob. sat.-ground. ($r = 0.5$m) & $p_{\text{T,sat-ground}}$ & 0.33\\
Trans. prob. sat.-sat. ($r=0.15$m) & $p_{\text{T,sat-sat}}$ & 0.052 - 0.0007\\
Trans. prob. sat.-sat. ($r=0.5$m) & $p_{\text{T,sat-sat}}$ & 0.55-0.03\\
\hline
\end{tabular}
\caption{Assumed values of the parameters in the model. The assumed values are compatible with current or near-term technology. The parameters that are different from the ones shown in this table are indicated in the figures. The collection efficiency is defined as $\eta_{\text{coll}}=p\eta_{\text{cav,coll}}$, where $p=0.99$ is the probability of emitting in the cavity mode, and $\eta_{\text{cav,coll}}=0.5$ is the collection efficiency from the cavity to free space. 
The effective loss coherence times depends on the quality of the vacuum in the atomic memories and is assumed to be higher for ground stations than on the satellites. The values of the transmission probability between satellites shown are the ones for the shortest distance between ground stations ($\sim 1300$ km) and for the longest one($\sim 12000 $km) to provide a sense of the relevant regime. }
\label{table:tab_1}
\end{table*}

Rather than using the rate analysis described above to estimate the achievable rate and fidelity for a fixed number of atoms, $N_{\text{mem}}$, we instead estimate the number of atoms needed to achieve a certain target rate and fidelity. We do this to get a sense of the multiplexing capabilities needed to reach fidelities above 90\% compatible with secret key distillation~\cite{gisin2002,Xu2020} and rates above 10Hz (100 Hz) such that at least 10 (100) entangled pairs could be distributed within the second long memory time that we assume for the ground state memories. Since the satellites are moving in orbit, the rate is estimated as the average rate during the time-window where ground-satellite communication is possible. This window is about 5 min for the satellite height of 500km considered here.  

In Fig.~\ref{fig:fig_3} we plot the necessary number of memory modes per repeater node, to reach a certain rate of entanglement between the ground stations, in this case $10$ Hz or $100$ Hz, and a fidelity $\geq0.9$ of the final Bell pair as a function of distance assuming a chain of five satellites. The parameters considered in the simulation are shown in Table~\ref{table:tab_1}. 

From the comparison of Figures~\ref{fig:fig_3} a) and b), it is seen that increasing the radius of the satellite mirrors relaxes the multiplexing requirements up to two orders of magnitude. While the Micius-like satellite requires more than $1000$ memory modes to get a rate of 100 Hz for a distance of 3500km between ground stations, the upgraded satellite only requires $\sim200$ memories.
Additionally, the upgraded satellite only requires around $1000$ memory modes to reach a rate of $100$ Hz over distances $\geq8000$ km. This is an order of magnitude less than found in optical BSM architectures for similar satellite characteristics~\cite{Liorni_2021}. 

We chose to target a final fidelity of $\geq0.9$ since this allows for direct extraction of secure encryption keys through quantum key distribution~\cite{gisin2002,Xu2020}. If higher fidelity entanglement is required, this can, in principle be achieved through entanglement purification~\cite{Victoria2023}. Increasing the rate from the targeted 10Hz or 100Hz can be achieved by increasing the amount of multiplexing. Increasing the number of memory modes in Fig~\ref{fig:fig_3} by a certain factor will result in roughly the same increase in rate.

As seen from table~\ref{table:tab_1}, quantum operations with \%-error level is sufficient  to reach a final fidelity $\geq0.9$. Lower error budgets in the quantum hardware will naturally increase the performance of the repeater. While errors at the 0.1\% level are within reach of current neutral atom based hardware, entanglement purification techniques could also be employed to leverage the effect of non-perfect operations at the expense of a slower distribution rate.   

\section{Conclusion}

In summary, we have proposed a satellite-assisted quantum repeater architecture based on individually trapped alkali atoms, which has several desirable features for functioning as quantum payloads. In particular, the ability to perform nearly deterministic BSMs with Rydberg-mediated two-qubit gates significantly lowers the amount of multiplexing needed to establish entanglement at continental distances compared to protocols based on probabilistic linear optics BSMs~\cite{Liorni_2021}. Additionally, the use of atoms as single photon sources circumvents the need for entangled photon sources and absorptive quantum memories.  

We developed a simple but accurate analytical model of the repeater architecture that allowed us to compute how expensive, in the sense of the number of memory modes per repeater station, it is to get high-fidelity entanglement at continental distances using the setup we propose. From this model, we estimated that a chain of 5 satellites with a $50$-centimeter radius telescope, enables high-fidelity ($F\geq 0.9$) entanglement at a rate of $100$ Hz with less than $200$ atoms per repeater station at a range of 1500 km.

We believe that our model can be readily adapted to provide first-order estimates of the performance for other satellite-assisted quantum repeater architectures based on heralded entanglement generation and entanglement swapping also with different quantum hardware. This is due to the characterization of the effect of the physical errors into simple lower-bound estimates of the fidelity of the final state. Notably our model allows for easy and fast simulation of long quantum repeater chains in contrast to Monte Carlo-based simulations~\cite{Wallnofer_2022} where the computational overhead increases rapidly with the length of the repeater chain.  

In addition to the promise of long-distance entanglement distribution, satellite-assisted quantum repeaters also open up new tools for more fundamental tests of nature. In particular, the atomic hardware considered here opens up new opportunities for quantum-enhanced sensor networks relevant for searches for topological dark matter~\cite{Belenchia2022}, global time-keeping~\cite{Komar2014}, and tests of the interplay between gravity and quantum mechanics~\cite{Bruschi2014}. \newline \newline
\textbf{Acknowledgements:} We acknowledge helpful discussions with Rudolf Saathof, Bob Dirks, and Gustavo Castro do Amaral. We acknowledge funding from the NWO Gravitation Program Quantum Software Consortium (Project QSC No. 024.003.037). J.B. acknowledges support from The AWS Quantum Discovery Fund at the Harvard Quantum Initiative. A. S. acknowledges the support of Danmarks Grundforskningsfond (DNRF 139, Hy-Q Center for Hybrid Quantum Networks). \newline \newline
\textbf{Data Availability:} The code and data of the results of this paper are openly available on 4TU.ResearchData: "Data underlying the publication "Satellite-assisted quantum communication with single photon sources and atomic memories", at \url{https://doi.org/10.4121/61635423-89d3-46cb-b9bc-bc06f158b9a5}, Ref.\cite{dataset}.

\begin{figure*}[htp]
    \centering
    \includegraphics[width=0.9\linewidth]{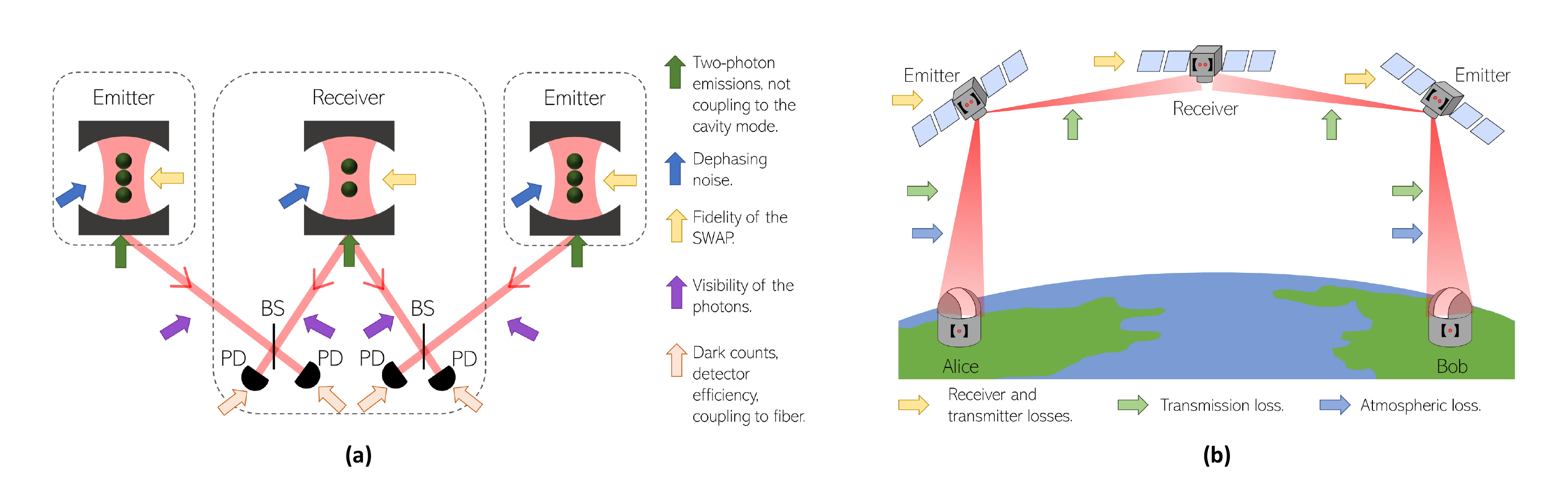}
    \caption{\textbf{Errors considered in the protocol. (a)} Schematic representation of the quantum setup used and possible errors that can happen. Two-photon emissions and imperfect coupling to the cavity mode are taken into account when modelling the atom-photon entanglement generation. Dark counts, detector efficiency and coupling to the fiber are included as imperfections in the Bell state measurements carried out in the receivers. Non-perfect visibility of the photons involved in the Bell state measurement is also included. Finally, we include dephasing of the atoms and imperfections of the SWAP operation. \textbf{(b)} Schematic representation of the transmission losses. In the satellite-satellite link, we take into account transmission losses due to the propagation of the Gaussian beam and the pointing jitter of the emitter. In the satellite-ground station link, we also consider the effect of the atmosphere, i.e. the Rayleigh scattering due to molecules. Additionally, losses originated in the terminals are also added to the model.}
    \label{fig:fig_5}
\end{figure*}

\appendix
\section{Modelling of Errors} \label{app:A}

In this appendix, we provide more details on our model of hardware imperfections and transmission loss in the satellite repeater chain.  Specifically, we describe how we model two-photon emission errors, dark counts, quantum memory dephasing, and atom loss. Finally, we describe our model of the optical transmission including beam divergence, pointing jitter, and atmospheric absorption. Additional details can be found in the supplementary material~\cite{SM}.  

In the first step of the protocol, for establishing entanglement between the spin of the emitter and the photon, we take into account two-photon emissions and that the photons may not be emitted into the cavity mode. The collected photons are sent to the closest satellite or to the ground station for the spin-spin entanglement generation. The photons will reach their destination with probability $p_\text{T}$, computed using the optical link budget explained later in this section. These free space and atmospheric losses are modelled as a fictitious beam splitter with transmission $p_\text{T}$. The transmission of a single photon is modelled as follows:
\begin{equation}
    \hat{a}^{\dagger}_{\text{ph}}|0\rangle_{\text{ph}}|0\rangle_{\text{E}} \rightarrow \sqrt{p_\text{T}}\hat{a}^{\dagger}_{\text{ph}}|0\rangle_{\text{ph}}|0\rangle_{\text{E}} + \sqrt{1-p_\text{T}}\hat{a}^{\dagger}_{\text{E}}|0\rangle_{ph}|0\rangle_{\text{E}},
    \label{eq:transmission_prob_BS}
\end{equation}
where $\hat{a}^{\dagger}_{\text{ph}}$ is the creation operator for the collected mode while $\hat{a}^{\dagger}_{\text{E}}$ is the creation of a non-collected (environment) photon. Once the photon reaches the receiver, a photonic Bell state measurement is performed between the emitted photon and a photon emitted from an atom in the receiver system. The two photons may not be perfectly indistinguishable and the efficiency of the single photon detectors and the coupling from free space to fiber is not perfect. These factors are added to the model as the visibility of the photons, the detector and the coupling efficiency. Moreover, dark counts can generate a "click" in the detectors, when there is no photon to measure, which is considered in the dark count probability. The photonic Bell state measurement is modelled as a $50/50$ beam splitter where just single photon emissions are considered indistinguishable leading to interference terms. In other words, we do not consider interference between photons from two-photon emissions since these will likely be emitted with very different temporal envelopes resulting in negligible interference effects. As a result of the above imperfections, the final atom-atom entangled states are captured by probabilistic mixtures of the ideal Bell state, a dephased state, and a non-specified garbage state with fidelity zero (which includes e.g. the $|00\rangle$ and $|11\rangle$ states) as specified in the supplemental material~\cite{SM}.

The entanglement is "stored" in the hyperfine spin of the Rubidium atoms but decoheres with time. We model this as a dephasing channel, where the coherence (off-diagonal terms of the density matrix) decayes exponentially with time as $\text{exp}(-t/\tau_{c})$, where $\tau_c$ is the coherence time of the atoms (assumed different for ground and satellites as detailed in Tab.~\ref{table:tab_1}). In addition, we assume that the atoms can get lost due to imperfect vacuum. We model this by assuming that with probability $\text{exp}(-t/T_{\text{loss}})$ an atom is lost after a time $t$, where $T_{\text{loss}}$ is an effective loss time (also assumed different for satellite and ground).  

After entanglement between elementary links is successfully achieved, we carry out the SWAP in all the satellites of the chain. To do so, the tweezers move the atoms that have established entanglement with their counterparts at the end of their respective elementary links to the top positions of the cavity, such that we know where the successfully entangled atoms are. The imperfections of the entanglement swap are modelled as described in the main text.

As shown in Fig.~\ref{fig:fig_5}b), we include a number of imperfections and loss for the optical link budget. The free space propagation losses are computed considering a fundamental Gaussian beam under transmitter pointing jitter. The latter is assumed to be described by a radially varying Rayleigh probability distribution function, which is the result of the combination of centered Gaussian probability density functions in both the vertical and the horizontal directions. The joint Gaussian probability density function of the pointing jitter is given by
\begin{equation}
    f(x_0,y_0;\sigma) = \dfrac {1}{2\pi\sigma^2 }\exp \left(-\frac{x_0^2+y_0^2}{2\sigma^2}\right)
\end{equation}
where $x_0$ and $y_0$ are the coordinates of the pointing error in the receiver's aperture plane and $\sigma$ is the pointing jitter's standard deviation for both directions. The effect of pointing errors is that the center of the Gaussian beam profile is not aligned with the center of the receiver. In the regime where the width $\sigma$, of the pointing jitter is small compared to the beam waist at the receiver, $w$, the effect of pointing jitter will be small and the transmission will be dominated by beam divergence. This allows us to use the binomial distribution approximation of the number of success full pairs in an elementary link used in our model (see Eq.~(\ref{eq:binom})) using the average transmission probability wrt. to the pointing jitter. From numerical Monte-Carlo simulations, we find that for $w/\sigma\geq10$, the binomial distribution approximation is good (see the supplemental material for further details~\cite{SM}). For realistic pointing errors and beam focusing characteristics~\cite{carrasco-casado_prototype_2022}, this requirement is  fulfilled~\cite{SM}.    
In satellite-to-ground links, the atmospheric effects have been included in the model using simulated atmospheric transmission from Ref.~\cite{giggenbach_atmospheric_2022} and assuming that beam wandering is dominated by transmitter pointing jitter.

\bibliography{biblio.bib}

\end{document}